\documentclass[aps,prd,amsmath,amssymb,showpacs]{revtex4}
\usepackage{graphics,epsfig,epsf,psfrag}
\usepackage{bm}
\usepackage{color}
\usepackage{amsmath}
\usepackage{amssymb}
\usepackage{amsfonts}
\usepackage{keyval,graphicx}
\usepackage{textcomp,wasysym}
\newcommand{\non}{\nonumber}

\newcommand{\gev}{~\mathrm{GeV}}
\newcommand{\mev}{~\mathrm{MeV}}

\newcommand{\y}{Y(4260)}

\newcommand{\DD}{\bar {D} D_1}
\newcommand{\fullprocess}{e^+e^-\to \chi_{c0}\omega}

\begin{document}

\title{Cross section line shape of $e^+e^-\to\chi_{c0}\omega$ around the $Y(4260)$ mass region }

\author{Martin Cleven$^{1,2}$\footnote{{\it E-mail address:} cleven@fqa.ub.edu},
        and Qiang Zhao$^{1,3}$\footnote{{\it E-mail address:} zhaoq@ihep.ac.cn}}

\affiliation{$^1$  Institute of High Energy Physics and Theoretical Physics Center for Science Facilities,\\
         Chinese Academy of Sciences, Beijing 100049, China}
\affiliation{$^2$ Departament de Fisica Quantica i Astrofisica and Institut de Ciencies del Cosmos\\
        Universitat de Barcelona, 08028-Barcelona, Spain}
\affiliation{$^3$ Synergetic Innovation Center for Quantum Effects and Applications (SICQEA), Hunan Normal
University, Changsha 410081, China}

\begin{abstract}
We demonstrate that the recent measurement of the cross section line shapes of $\fullprocess$
can be naturally explained by the molecular picture for $\y$ where the $\y$ is treated as a hadronic molecule dominated by $\bar{D}D_1(2420)+c.c.$ This result is consistent with properties extracted for $Y(4260)$ as the $\bar{D}D_1(2420)+c.c.$ molecular state in other reactions such as $e^+e^-\to J/\psi\pi\pi$, $h_c\pi\pi$, $\bar D D^*\pi +c.c.$, and $\gamma X(3872)$.
\end{abstract}



\date{\today}
\maketitle
\section{Introduction}
Ever since its discovery in 2005 by BABAR Collaboration~\cite{Aubert:2005rm} the true nature of $\y$ has attracted a lot of attention from the hadron physics community. Its mass, $M_Y=(4251\pm 9)\mev$, was first extracted from the $J/\psi\pi\pi$ invariant mass spectrum by BABAR with a width of $\Gamma=(120\pm 12)\mev$~\cite{Aubert:2005rm}.
With the relatively well-established quark model states $\psi(2D)$ and $\psi(4S)$ as successive states in the energy levels for the vector charmonium spectrum, the presence of $Y(4260)$ between those two states have indicated some unusual origin and raised questions about the underlying dynamics. The other surprising feature of $Y(4260)$ is its ``absence" from the open charm decay channels, such as $D\bar{D}$, $D\bar{D}^*+c.c.$ and $D^*\bar{D}^*$. Namely, there are no obvious peak structures around the $Y(4260)$ mass region in $e^+e^-$ annihilations into open charm final states.

Following these observations there have been various theoretical interpretations for the nature of $\y$ in the literature. In Refs.~\cite{Zhu:2005hp,Kou:2005gt,Close:2005iz}, $\y$ was proposed to be the charmonium hybrid candidate in the vector sector if its decay would be dominated by the $J/\psi\pi\pi$ channel. Another interesting feature is that the Lattice QCD simulations favor a charmonium hybrid located in this region. However, more experimental measurements~\cite{Yuan:2013lma} impose challenging questions on this scenario~\cite{Berwein:2015vca}. Voloshin {\it et al.} proposed the hadro-quarkonium picture trying to understand the peculiar decay behavior of $\y$~\cite{Voloshin:2007dx,Dubynskiy:2008mq}. Tetraquark solutions were also studied in the literature taking into account diquark correlations~\cite{Maiani:2005pe}. In Refs.~\cite{LlanesEstrada:2005hz,He:2014xna} $\y$ was assigned to be the conventional charmonium state $\psi(4S)$.
In Refs.~\cite{Barnes:2007xu,Li:2009ad,Segovia:2008zz} the coupled-channel effects in the charmonium spectrum were proposed as an explanation for the observed peaks that cannot be assigned to conventional charmonia. 
It was also proposed that $\y$ could be hadronic molecules composed of $\bar{D}D_1(2420)+c.c.$~\cite{Close:2008hv,Ding:2008gr}, $\chi_{c0}\omega$~\cite{Dai:2012pb}, or $J/\psi K\bar{K}$~\cite{MartinezTorres:2009xb}.
Among these molecular solutions it is difficult to accommodate the $J/\psi K\bar{K}$ picture within the updated experimental measurements in various processes, while $\bar{D}D_1(2420)+c.c.$ and $\chi_{c0}\omega$ bound state interpretations are attractive to be confronted by experimental data for both hidden charm decay channels, i.e. $e^+e^-\to J/\psi\pi\pi$, $h_c\pi\pi$, $\chi_{c0}\omega$, or open charm decays into $D\bar{D}^*\pi+c.c.$

Amid the growing interest the recent Lattice QCD (LQCD) simulations make the interpretation of $\y$ a nontrivial task.
The TWQCD Collaboration~\cite{Chiu:2005ey} finds a resonance with mass around $(4.238\pm0.031)\gev$ for the molecular operator which can be identified as the $\y$ state while the hybrid operator leads to much higher masses above 4.2 GeV.
In contrast, two other LQCD groups find quite different results.
The Hadron Spectrum Collaboration~\cite{Liu:2012ze} calculates the charmonium spectrum at the pion mass of $400\mev$ and obtains the lightest vector charmonium hybrid around 4.2 GeV which can be possibly related to $\y$ as a vector hybrid meson. Chen et al.~\cite{Chen:2016ejo} also see a charmonium hybrid vector with a mass of $(4.33\pm0.02)\gev$. Note that in Refs.~\cite{Liu:2012ze,Chen:2016ejo} the large operator overlapping comes from the configuration that the $\bar{c}c$ component has $J^{PC}=0^{-+}$ and the gluon field with $J^{PC}=1^{+-}$. It implies the leading suppression of the hybrid production in $e^+e^-$ annihilations in the HQSS limit. The leptonic decay width of the hybrid vector is estimated to be narrow~\cite{Close:2005iz}, and the estimate of LQCD is about 40 eV~\cite{Chen:2016ejo} as a feature of the hybrid scenario.

Triggered by the recent observation of charged charmonium candidates $Z_c(3900)$ and $Z_c(4020)$ by BESIII, in a series of recent works~\cite{Wang:2013cya,Guo:2013nza,Cleven:2013mka,Wang:2013kra,Qin:2016spb} it was demonstrated that the present experimental measurements provide strong evidences for the $\y$ being a  $\bar{D}D_1(2420)+c.c.$ molecular state. The interpretation agrees with the current data for the decay channels $\y\to J/\psi\pi\pi$, $\y\to h_c\pi\pi$ and, crucially, $\bar D D^*\pi +c.c.$  The latter channel is expected to contribute to a large branching fraction for $\y$. However due to the dynamics involved in the open threshold regime, a clear, Breit-Wigner-like cross section line shape is not expected. This phenomenon is taken as a support of the hadronic molecule scenario~\cite{Cleven:2013mka,Qin:2016spb} and turns out to be coincide with the experimental measurement of Belle~\cite{Pakhlova:2009jv}. It should be noted that so far none of those observed enhancement structures in the mass region of $\y$ appears to be consistent with a single Breit-Wigner distribution. However, this is exactly what should happen in the molecular picture. The cross section lineshapes in exclusive channels can be very different due to the near-threshold interactions, while the pole position of the state keeps the same in all the channels. Because of this, it could be very misleading to interpret these structures as ``different states" when a strongly coupled $S$-wave threshold is present.

The new experimental data provided by the BESIII Collaboration~\cite{Ablikim:2014qwy,Ablikim:2015uix} allow the existing interpretations of the $\y$ structure to be tested in the decay channel of $\chi_{cJ}\omega$. The data show a resonance-like structure with a peak position at $4.220\mev$ in $\chi_{c0}\omega$ with the absence of the same structure in $\chi_{c1, c2}\omega$.
There are several ideas of how to treat this new structure.
Li and Voloshin~\cite{Li:2014jja} explain the process through the known resonance $Y(4160)$.
Faccini {\it et al.}~\cite{Faccini:2014pma} interpret the structure in the $\chi_{c0}\omega$ channel using Breit-Wigner functions and masses for a tetraquark state determined in Ref.~\cite{Maiani:2014aja} to describe the data.

In this work, we continue to explore the consequence of the  $\bar{D}D_1(2420)+c.c.$ molecular scenario from Refs.~\cite{Wang:2013cya,Cleven:2013mka,Qin:2016spb} in the $e^+e^-\to \omega\chi_{cJ}$ channel. The idea is to understand the relation between the  $\bar{D}D_1(2420)+c.c.$ and $\chi_{c0}\omega$ thresholds and their impact on the cross section line shape of $e^+e^-\to \omega\chi_{cJ}$. We try to constrain the inevitable new parameters and make predictions based on the $\bar{D}D_1(2420)+c.c.$ molecular scenario. This should be a challenging test of what we learned about $\y$ in other channels.

This work is structured as follows: In Sec.~\ref{sec:framework} we will present cornerstones of the previous studies of $\y$ in the molecular interpretation together with the interactions of the charmed mesons that drive the decay in this picture. Section~\ref{sec:results} contains the results for the line shape as well the estimate of the branching fraction. A brief summary is given in Sec.~\ref{sec:summary}.
\section{Framework}\label{sec:framework}

In the picture of $\DD$ (from here on we will use the short-handed notation $\DD$ for $\bar{D}D_1(2420)+c.c.$) molecular state for $\y$, the Lagrangian for the $\y$ couplings to $\DD$ is expressed as the following~\cite{Wang:2013cya,Cleven:2013mka,Qin:2016spb}:
\begin{eqnarray}\label{lagrangian-YD1D}
\mathcal{L}_Y=\frac{y^{\text{bare}}}{\sqrt{2}}(\bar{D}_a^\dagger Y^iD^{i\dagger}_{1a}-\bar D^{i\dagger}_{1a} Y^iD^\dagger_a)
+g_c[(D^i_{1a}\bar{D}_a)^\dagger(D^i_{1b}\bar{D}_b)+(D_{a}\bar{D}^i_{1a})^\dagger(D_{b}\bar{D}^i_{1b})]+H.c. , \non
\end{eqnarray}
where $y^{\text{bare}}$ stands for the bare coupling between the bare $\y$ state and $\DD$ while $g_c$ describes the non-resonant contact interaction for $\DD\to \DD$.
The bare $\y$ is then dressed by the $\DD$ rescatterings. The detailed deductions are referred to Refs.~\cite{Wang:2013cya,Cleven:2013mka,Qin:2016spb}. In the following we will only present the results relevant for the calculations presented here.
The dressed non-relativistic propagator of the $Y(4260)$ can be expressed as
\begin{eqnarray}\label{propagator-full}
\mathcal{G}_Y(E)&=&\frac{1}{2m_Y}\frac{iZ}{E-m_Y-Z\widetilde{\Sigma}_1(E)+i\Gamma^{\text{non}-\bar{D}D_1}/2},
\end{eqnarray}
where $m_Y$ is the renormalized mass of $\y$ and $\Gamma^{\text{non}-\bar{D}D_1}$  accounts for contributions from decay channels other than the $\DD$~\cite{Cleven:2013mka,Qin:2016spb}. Both are fitted to experimental data for the cross section lineshapes of $e^+e^-\to J/\psi\pi\pi$ and $h_c\pi\pi$ and found to be $m_Y=4.217\pm0.002$ GeV and $\Gamma^{\text{non}-\bar{D}D_1}=0.056\pm0.003$ GeV~\cite{Cleven:2013mka}.
The self-energy  stems from the sum of the infinite bubble loops in the $\bar{D}D_1$ rescatterings.
Using dimensional regularization with the $\overline{\text{MS}}$ subtraction scheme, one finds
\begin{eqnarray}
\widetilde{\Sigma_1}(E)= \quad\Sigma_1(E)-{\rm Re}(\Sigma_1(m_Y))-(E-m_Y){\rm Re}(\Sigma_1'(m_Y)),
\end{eqnarray}
with
\begin{equation}
\Sigma_1(E)\equiv\Sigma_{\bar{D}D_1}(E)((y^\text{bare})^2-4(E-m_0)g_1),
\end{equation}
and
\begin{equation}
\Sigma_{\bar{D}D_1}(E) \equiv \frac{\mu}{8\pi}\sqrt{2\mu(m_D+m_{D_1}-E)-i\mu\Gamma_{D_1}} \ ,
\end{equation}
where $\mu=m_D m_{D_1}/(m_D+m_{D_1})$ is the reduced mass. The bare couplings in Eq.~(\ref{lagrangian-YD1D}) get renormalized to the effective coupling
\begin{equation}
 y_\mathrm{eff}\equiv \sqrt{Z}y^{\text{bare}}=(3.94\pm0.04)\gev^{-1/2},
\end{equation}
and $g_c=(29.50 \pm 0.47) \gev^{-2}$ is determined by the fitting.
The same renormalization constant $Z$ can be used to obtain the physical coupling of the $\y$ to a photon
\begin{eqnarray}
\mathcal{L}_{Y\gamma}&=&\frac{em^2_Y}{f_Y}Y_\mu A^\mu. \
\end{eqnarray}
The physical coupling was determined as $\frac{1}{f^\text{eff}_Y}\leq\sqrt{Z}\frac{1}{f_Y}=0.023\pm 0.004$ by Ref.~\cite{Qin:2016spb}.
\begin{figure*}
\centering
\includegraphics[width=0.4\linewidth]{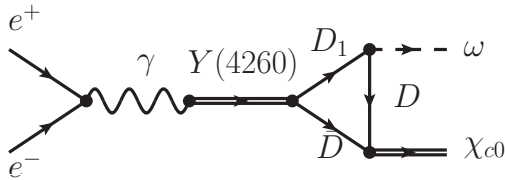}
\caption{The Feynman diagram for the leading loop transition for $e^+e^-\to \y\to\chi_{c0}\omega$. } \label{fig:feynman}
\end{figure*}


We use the well-established method of vector meson dominance (VMD) to estimate the coupling of the light vector meson $\omega$ to the charmed meson multiplets. Comprehensive studies and review of the VMD can be found in Refs.~\cite{Meissner:1987ge,O'Connell:1995wf} and references therein.
The relevant Lagrangian that couples the charmed and their anti-charmed partners to the $\omega$ meson in the non-relativistic effective field theory (NREFT) is given by
\begin{equation}\label{coupl-1}
 \mathcal L _{TH\omega}=\frac{c_\omega}2\left< T_b^i H_a^\dag - \bar {T}_b^i \bar {H}_a^\dag\right> \omega_{ab}^i \ ,
\end{equation}
where the charmed meson fields $H_a$ for the pseudoscalar and vector meson spin doublets are
\begin{equation}
 H_a=V_a\cdot \sigma + P_a \ ,
\end{equation}
and the $D_1$ and its spin partner $D_2$ in the HQSS limit are expressed as
\begin{equation}
 T_a^i = D_{2a}^{ij}\sigma^j + \sqrt{2/3}D_{1a}^i + i \sqrt{1/6} \varepsilon^{ijk}D_{1a}^j\sigma^k \ .
\end{equation}
The coupling in Eq.~(\ref{coupl-1}) can be determined as $c_\omega = 1.62\pm 0.01$. A more detailed derivation can be found in App.~\ref{app:vmd}.

Finally, the coupling of  charmed mesons to $P$-wave charmonia reads
\begin{eqnarray}
{\cal L}_{\chi} &=& i\frac{g_1}{2}\left<\chi^{\dag i}H_a\sigma^i\bar H_a\right> + {\rm H.c.}.
\end{eqnarray}
In a recent work~\cite{Cao:2016xqo} the radiative decays of vector charmonia to $\chi_{cJ}$ were studied and the coupling $g_1$ was determined as $g_1 = (4.99\pm 0.92)\gev^{-1/2}$ and agreed with previous calculations using QCD sum rules.
\section{Results and Discussions}\label{sec:results}

The matrix element for the process $\fullprocess$ as shown  in Fig.~\ref{fig:feynman} reads
\begin{eqnarray}
 \mathcal M = \bar v(p_1)(-ie \gamma^\mu) u(p_2) \frac{i}{E^2}\frac{iem_Y^2}{f_Y^2}G_Y(E)  (-\sqrt 8)g_1 g_Yc_\omega \mathcal{I}^0 \sqrt{m_Y m_\chi}\varepsilon^i(\omega)
\end{eqnarray}
where $\mathcal {I}^0 =\mathcal {I}^0 (m_{D_1},m_D,m_{D};E,m_{\chi_{c0}},m_\omega)$ is the scalar three-point loop as derived in Ref.~\cite{Guo:2013nza}.

In principle, all couplings in this matrix element are known, although in all cases they come with sizable uncertainties.
In order to compromise the uncertainties arising from the couplings, we choose to fix the couplings using their central values and at the same  time introduce a dimensionless overall constant $\mathcal C$ to measure the overall deviations of the calculation compared with the experimental data. Namely, only parameter $\mathcal C$ will be fixed by fitting the experimental data.

\begin{figure}
\centering
 \includegraphics[width=0.8\linewidth]{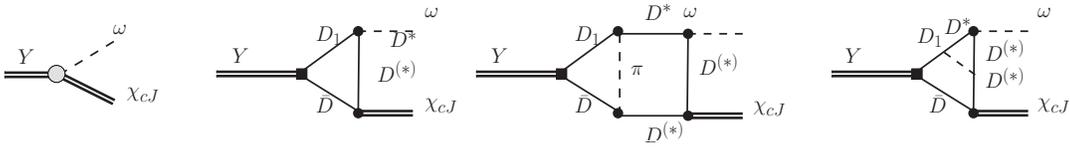}
 \caption{Different contributions to $\y\to\chi_{cJ}\omega$.}
 \label{fig:higher-loop}
\end{figure}
In the framework of NREFT we expect that higher loop contributions are relatively suppressed in comparison with the one-loop amplitude.
Following the power counting scheme of Ref.~\cite{Guo:2010ak} and assuming that the contact term for $\y\to\chi_{cJ}\omega$ scales as ${\cal O}(1)$ we estimate that the loop amplitude of Fig.~\ref{fig:feynman} scales as $v^5(v^{-2})^3\sim v^{-1}$. With $v\sim 0.1$ the loop contribution will be enhanced compared to the contact term.
In Figs.~\ref{fig:higher-loop} (c) and (d) the higher loops that include an internal light meson line, e.g. a pion exchange, scale as $(v^5)^2(v^{-2})^6 v^2 v^2\sim v^2$ and will be strongly suppressed.
In contrast, the amplitude of Fig.~\ref{fig:higher-loop} (e) scales as $(v^5)^2(v^{-2})^6 v^2\sim 1$ which can be absorbed into the redefinition of the contact term coupling. Thus, we only consider the triangle loop in the calculation as a reasonable estimate of the transition rate for $\y\to\chi_{cJ}\omega$.

\begin{figure}
\centering
 \includegraphics[width=0.7\linewidth]{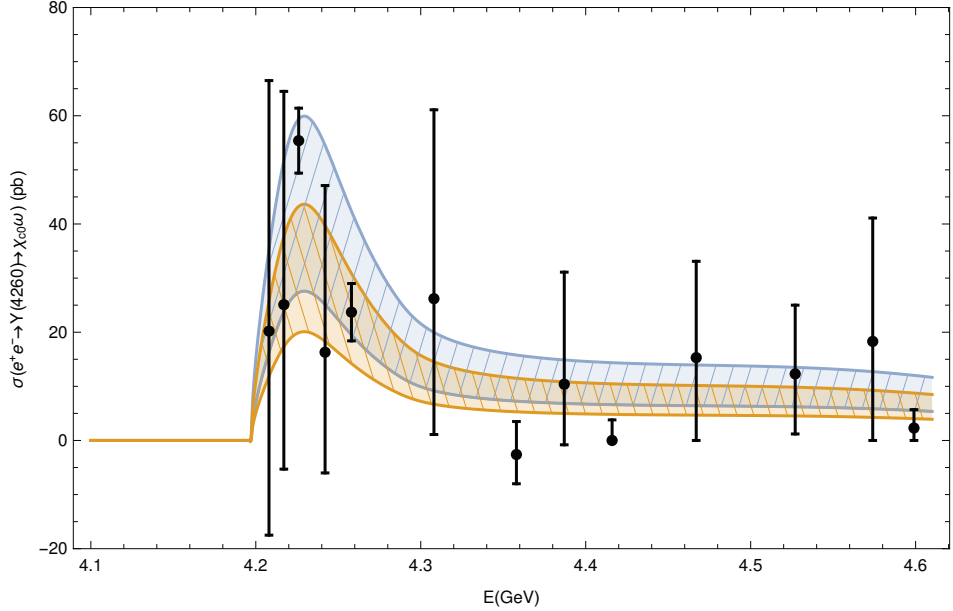}
 \caption{Cross section lineshapes fitted to the BESIII experimental data~\cite{Ablikim:2014qwy}.
 The upper band (blue) only fits the first seven points, while the lower band (brown) uses all data. }
 \label{fig:lineshape}
\end{figure}

Before we proceed to the numerical results, it is interesting to have a closer look at the experimental data for $\fullprocess$, shown by the black dots in Fig.~\ref{fig:lineshape}. There appears an immediate increase of the cross section when the phase space opens. Although most of the data points still contain large errors those points located at the peak position and down-slope of the peak have much smaller errors which dictate the peaking around 4.22 GeV. The cross sections then seem to be vanishing around 4.35 GeV and flat away beyond 4.36 GeV. There are two challenges for any model studies: (i) Whether the threshold peak can be reproduced? (ii) Whether the cross section magnitude can be understood?

We take two steps to fit the experimental data taking into account that only one parameter is present in the amplitude. Firstly, we fit the data in the peak region. Namely, we only fit the first 7 data points. A reduced $\chi^2$ is found to be $\chi^2/N_{d.o.f}=13/7\sim2$ and the cross section is shown by the upper band in Fig.~\ref{fig:lineshape}.
We then fit all the data simultaneously and the reduced $\chi^2$ becomes $\chi^2/N_{d.o.f}=3$.
The corresponding cross sections are presented by the lower band in Fig.~\ref{fig:lineshape}.

Comparing these two fits one can see that it is the data near threshold that determine the behavior of the fitted cross sections and these two fitting results have reasonably described the threshold peaking. The fitted values of $\mathcal C$ are 0.11 and 0.09 for the first 7 and all data points, respectively. With $\mathcal C<1$ it suggests that the product of the central values for the coupling constants $g_Yg_1 c_\omega$ has overestimated the experimental data. Considering that the cross section is proportional to $(g_Yg_1 c_\omega)^2$, the overall uncertainty is about $30\%$ and can be regarded as acceptable since it is still within the uncertainties of these couplings.
On top of the uncertainties stemming from the coupling constants we need to consider the theoretical uncertainty that is associated with the NREFT approach.
But as estimated by the approximate power counting for Fig.~\ref{fig:higher-loop}, the higher loop contributions to the uncertainties are much smaller than the uncertainties from the coupling constants.

It should be noted that our calculations here only considered the molecular component for $\y$ which is dominant as shown in Ref.~\cite{Qin:2016spb}. We neglected the possible direct contributions for the small $c\bar{c}$ component coupling to $\chi_{c0}\omega$. The empirical reason is that the $c\bar{c}$ component of $\y$ is small in our picture. If this component can produce the threshold enhancement via the tree level coupling, one would also expect that there will be threshold enhancement for $e^+e^-\to \chi_{c1,c2}\omega$ due to the HQSS. Since there is no obvious threshold enhancement present in these two channels~\cite{Ablikim:2014qwy,Ablikim:2015uix}, it is reasonable to neglect the possible direct coupling of the small $c\bar{c}$ spin-1 component to the final $\chi_{c0}\omega$.
Taking into account that $\mathcal C<1$ has implied the dominance of the molecular component in the transition, we regard the results are consistent with treating $\y$ as the $\DD$ molecule.

One also notices that the data in Fig.~\ref{fig:lineshape} suggest that there are no obvious couplings of a conventional charmonium state to $\chi_{c0}\omega$ beyond the threshold region. This means that even for the conventional vector charmonium states the couplings to the $\chi_{c0}\omega$ are rather small due to HQSS.
The molecular scenario, however, allows for  a sizable coupling to this final state through the loop mechanism.

In the $\DD$ molecular scenario, given that the locations of the $\chi_{c1}\omega$ and $\chi_{c2}\omega$ thresholds, $4.29\gev$ and $4.34\gev$, respectively, are much higher than the pole position of $\y$, it suggests that the influence of $\y$ in these two channels will only be marginal which is consistent with the experimental observations~\cite{Ablikim:2014qwy,Ablikim:2015uix}.

One cannot deny that the absence of structures beyond the $\y$ mass may be caused by more complicated coupled-channel interferences when more thresholds, i.e. the $\DD^*$ and $D_2D^*$ thresholds, and more charmonium states can contribute. Detailed answer to this question should rely on more comprehensive calculations which, however, has gone beyond the scope of this work. In the energy region close to the first narrow $S$-wave threshold $\DD$ the transition mechanism can still be reasonably approximated by the single-channel problem.

It should also be pointed out that the cross section around the peak is about 50 pb which is the same order as that for the $J/\psi\pi\pi$ channel, but nearly one order of magnitude smaller than that for the $D\bar{D}^*\pi+c.c.$ channel. The interesting observation is that the $\DD$ molecular picture can easily understand the relatively suppressed cross sections for the $\y\to J/\psi\pi\pi$ and $\chi_{c0}\omega$ channel due to the loop transitions, and the non-trivial lineshape for $D\bar{D}^*\pi+c.c.$ as shown in Refs.~\cite{Cleven:2013mka,Qin:2016spb}.

Our investigation of the cross section lineshape in Fig.~\ref{fig:lineshape} suggests that the cross sections for $e^+e^-\to\chi_{c0}\omega$ in the threshold region is dominated by the production of $\y$ as the $\DD$ molecule. Thus, we can estimate the partial decay width for $\y\to \chi_{c0}\omega$ via the triangle diagram. This yields
\begin{equation}\label{Eq:WidthY}
 \Gamma(Y\to\chi_{c0}\omega) = \frac{1}{3\pi}\frac{|\vec q|}{m_Y} \mathcal {C}^2\!g_1^2g_Y^2c_\omega^2 m_\chi|\mathcal{I}^0|^2
\end{equation}
In order to compare with the experimental measurement, we take into account the width effects of the $\y$ to extract the partial width. To do so, we calculate the partial width with the integration over the spectral function
\begin{eqnarray}
 \bar{ \Gamma}(Y(4260)\to \chi_{c0}\omega) = \frac1{W}\int_{(m_Y-2\Gamma_Y)}^{(m_Y+2\Gamma_Y)}
  \!\!dE\, \Gamma(Y\to\chi_{c0}\omega) \frac1{\pi}{\rm Im}\left(\frac{-1}{E-m_Y+i\Gamma_Y/2}\right)
\end{eqnarray}
where the normalization is defined as
\begin{eqnarray}
  W &=&
  \int_{(m_Y-2\Gamma_Y)}^{(m_Y+2\Gamma_Y)}
  \!\!dE\,  \frac1{\pi}{\rm Im}\left(\frac{-1}{E-m_Y+i\Gamma_Y/2}\right).
\end{eqnarray}
For an exclusive process the total width $\Gamma_Y$ in the above equation is an input value.
Notice that we use the results from our model, $m_Y=4.22\gev$ and $\Gamma_Y=0.075\gev$ rather than the PDG values obtained from fits with Breit-Wigner distributions.
We find $\bar{ \Gamma}(Y(4260)\to \chi_{c0}\omega) = (1.4\pm 0.3)\mev$, compared to $(1.2\pm0.2)\mev$ without taking the finite width into account.
This is consistent with Ref.~\cite{Qin:2016spb} where the partial width of 1.6~MeV was extracted without finite width corrections.
Compared to the dominant decay channel of $\y\to D\bar{D}^*\pi+c.c.$~\cite{Qin:2016spb}, the decay channel $\y\to \chi_{c0}\omega$ is suppressed by more than one order of magnitude.

When we compare our $\DD$  with the $\chi_{c0}\omega$ molecular picture proposed by Ref.~\cite{Dai:2012pb},
we notice several crucial differences for these two interpretations.

Firstly, the cross section magnitude of $e^+e^-\to D\bar{D}^*\pi+c.c.$ needs to be measured precisely. Our $\DD$ molecular scenario predicts a rather large cross section for $e^+e^-\to \y\to D\bar{D}^*\pi+c.c.$ and a non-trivial cross section lineshape in the vicinity of $\y$ which can be examined by forthcoming experimental measurements.

Secondly,  the leptonic decay width for $\y\to e^+e^-$ is predicted very differently in these two scenarios.
The magnitude of the leptonic decay width determines how the strong decay widths sum up to the total width.
Smaller leptonic decay width means that the strong decay widths will be relatively enhanced and vice versa.
Because of this the measurement of cross sections for various decay channels is useful for disentangling the transition mechanism by comparing the relative strong decay strengths.
In our model the dominant decay width is the $D\bar{D}^*\pi+c.c.$ channel and the molecular nature of $\y$ leads to a relatively small partial widths for other channels such as $J/\psi\pi\pi$, $h_c\pi\pi$ and $\chi_{c0}\omega$.
Meanwhile, the resonance parameters cannot be fitted by a simple Breit-Wigner, and the extracted total width of $(73.9\pm 4.5)$ MeV is smaller than the Breit-Wigner width of about 120 MeV.
Thus, it allows the partial width of $\y\to e^+e^-$ to be at the order of about 500 eV~\cite{Qin:2016spb}.
For the $\chi_{c0}\omega$ molecule, the predicted leptonic decay width is only about 23 eV~\cite{Dai:2012pb}.
This is because the partial widths for the $J/\psi\pi\pi$ and $\chi_{c0}\omega$ channels have been fitted to be large in order to account for the total width of about 100 MeV, and no contributions from the open charm decay channel are included.
We also mention that the LQCD also predicts a very small leptonic decay width of $< 40$ eV for a hybrid vector charmonium state~\cite{Chen:2016ejo}.
Therefore, the experimental extraction of the leptonic decay width for $\y\to e^+e^-$ is important for distinguishing the $D_1D$ molecule solution from the $\chi_{c0}\omega$ molecule and hybrid scenario.

Thirdly, these two scenarios should have different partial decay widths for $\y\to \gamma X(3872)$. This quantity was predicted to be sizeable in Ref.~\cite{Guo:2013nza} as a consequence of the $\DD$ molecular scenario and confirmed by the BESIII measurement~\cite{Ablikim:2013dyn}. In contrast, the $\chi_{c0}\omega$ picture may also lead to a sizeable branching ratio to $\gamma J/\psi\omega$ via an internal E1 transition of $\chi_{c0}\to \gamma J/\psi$ and then break up the bound system. But it is unlikely to have a sizeable coupling to $X(3872)$ since it is subleading effect for $J/\psi\omega$ scattering to generate $X(3872)$.

\section{Summary}\label{sec:summary}
%
In this work, we demonstrate that the experimental data for $e^+e^-\to\chi_{c0}\omega$ in the threshold energy region provide important information for the underlying dynamics where the mysterious state $\y$ can be explained as the $\bar{D}D_1(2420)+c.c.$ molecule. The rescatterings of the $\bar{D}D_1(2420)+c.c.$ into the final state $\chi_{c0}\omega$ explain the much smaller cross sections for $e^+e^-\to\chi_{c0}\omega$ compared to the channel $e^+e^-\to D\bar{D}^*\pi+c.c.$ although the latter has a non-trivial cross section lineshape~\cite{Cleven:2013mka,Qin:2016spb}. We also emphasize that the observed peak position is located near the pole mass for $\y$ which has been studied in detail in Refs.~\cite{Cleven:2013mka,Qin:2016spb}. Thus, this new set of data can be accommodated consistently in the molecular picture for $\y$. We also note that the energy region near threshold involves much less contributing mechanisms which is ideal for us to clarify the role played by the $\bar{D}D_1(2420)+c.c.$ open charm. In contrast, the experimental data show that the cross sections for $e^+e^-\to\chi_{c0}\omega$ flat out beyond 4.36 GeV. Whether it is due to the interferences of multi-processes or absence of significant contributions from resonances, it needs further elaborate coupled-channel studies in the future.
%
\section{Acknowledgments}
%
Useful discussions with F.-K. Guo, C. Hanhart and Q. Wang are acknowledged. This work is supported, in part, by the National Natural Science Foundation of China (Grant Nos. 11425525, and 11521505), DFG and NSFC funds to the Sino-German CRC 110 ``Symmetries and the Emergence of Structure in QCD'' (NSFC Grant No. 11261130311), and National Key Basic Research Program of China under Contract No. 2015CB856700.
MC is also supported by the Chinese Academy of Sciences President's International Fellowship Initiative grant 2015PM006, the Spanish Ministerio de Economia y Competitividad (MINECO) under the project MDM-2014-0369 of ICCUB (Unidad de Excelencia 'Mar\'\i a de Maeztu'), and, with additional European FEDER funds, under the contract FIS2014-54762-P as well as support from the Ge\-ne\-ra\-li\-tat de Catalunya contract 2014SGR-401, and from the Spanish Excellence Network on Hadronic Physics FIS2014-57026-REDT.
\begin{appendix}
\section{Vector meson dominance}\label{app:vmd}
The derivation here is similar to one done by O'Connel {\it et al.} in~\cite{O'Connell:1995wf}. We have the Lagrangian that couples light vector mesons to a $TH$ pair
\begin{equation}
 \mathcal L _{TH\omega}=\frac{c_\omega}2\left< T_b^i H_a^\dag - \bar {T}_b^i \bar {H}_a^\dag\right> \omega_{ab}^i.
\end{equation}
and the Lagrangian for the same pair coupling to a photon
\begin{equation}
 \mathcal L _{TH\omega}=\frac12\left< T_b^i H_a^\dag - \bar {T}_b^i \bar {H}_a^\dag\right> E^i\left( e\frac{\beta_Q}{m_Q}Q'+e \beta_q Q_{ab}\right).
\end{equation}
Here $E^i$ is the electric field, $Q$ is the light quark charge matrix and $Q'$ the heavy quark charge. Quantity $\beta_Q$ is related to the Isgur-Wise function in the non-recoil approximation \cite{Korner:1992pz} and takes the ISGW value 0.584 from Ref.~\cite{Isgur:1990jf}. We also take the value $\beta_q = 0.634 \pm 0.034 \gev ^{-1}$ the same as that in Ref.~\cite{Korner:1992pz}.

According to~\cite{O'Connell:1995wf} the relation between the light current coupling $\beta_q$ and the coupling of the light vector mesons is in general given by
\begin{equation}
 \sum_V g_V \frac{1}{q^2-m_V^2}\frac{em_V^2}{f_V}\frac{1}{E_\gamma} = e_q.
\end{equation}
Applying this to the process $D_1^+\to D^+ \gamma$ for small momenta $q$ we find
\begin{equation}
 -\frac13 \sqrt{\frac23}\beta_q e = \frac{1}{\sqrt3}\frac{c_\omega}{E_\gamma} \left(\frac{e}{f_\omega}-\frac{e}{f_\rho}\right)
\end{equation}
Using the values $e/f_\rho = 6.12 \times 10^{-2}$ and $e/f_\omega = 1.78 \times 10^{-2}$ extracted from their dilepton decay widths we obtain  $c_\omega = 1.62\pm 0.01$.
\end{appendix}


\begin{thebibliography}{99}
\bibitem{Aubert:2005rm}
  B.~Aubert {\it et al.}  [BaBar Collaboration],
  Phys.\ Rev.\ Lett.\  {\bf 95}, 142001 (2005).


\bibitem{Zhu:2005hp}
  S.-L.~Zhu,
  Phys.\ Lett.\ B {\bf 625}, 212 (2005).

\bibitem{Kou:2005gt}
  E.~Kou and O.~Pene,
  Phys.\ Lett.\ B {\bf 631}, 164 (2005).

\bibitem{Close:2005iz}
  F.~E.~Close and P.~R.~Page,
  Phys.\ Lett.\ B {\bf 628}, 215 (2005).

\bibitem{Yuan:2013lma}
  C.~Z.~Yuan,
  arXiv:1310.0280 [hep-ex].

\bibitem{Berwein:2015vca}
  M.~Berwein, N.~Brambilla, J.~Tarr\`{u}s Castell\`{a}, and A.~Vairo,
  Phys.\ Rev.\ D {\bf 92} (2015) no.11,  114019
  doi:10.1103/PhysRevD.92.114019
  [arXiv:1510.04299 [hep-ph]].

\bibitem{Voloshin:2007dx}
  M.~B.~Voloshin,
  Prog.\ Part.\ Nucl.\ Phys.\  {\bf 61}, 455 (2008).

\bibitem{Dubynskiy:2008mq}
  S.~Dubynskiy and M.~B.~Voloshin,
  Phys.\ Lett.\ B {\bf 666}, 344 (2008).


\bibitem{Maiani:2005pe}
  L.~Maiani, V.~Riquer, F.~Piccinini and A.~D.~Polosa,
  Phys.\ Rev.\ D {\bf 72}, 031502 (2005)
  doi:10.1103/PhysRevD.72.031502
  [hep-ph/0507062].

\bibitem{LlanesEstrada:2005hz}
  F.~J.~Llanes-Estrada,
  Phys.\ Rev.\ D {\bf 72}, 031503 (2005).


\bibitem{He:2014xna}
  L.~P.~He, D.~Y.~Chen, X.~Liu and T.~Matsuki,
  Eur.\ Phys.\ J.\ C {\bf 74} (2014) no.12,  3208
  doi:10.1140/epjc/s10052-014-3208-5
  [arXiv:1405.3831 [hep-ph]].


\bibitem{Barnes:2007xu}
  T.~Barnes and E.~S.~Swanson,
  Phys.\ Rev.\ C {\bf 77} (2008) 055206
  doi:10.1103/PhysRevC.77.055206
  [arXiv:0711.2080 [hep-ph]].

\bibitem{Li:2009ad}
  B.~Q.~Li, C.~Meng and K.~T.~Chao,
  Phys.\ Rev.\ D {\bf 80} (2009) 014012
  doi:10.1103/PhysRevD.80.014012
  [arXiv:0904.4068 [hep-ph]].

\bibitem{Segovia:2008zz}
  J.~Segovia, A.~M.~Yasser, D.~R.~Entem and F.~Fernandez,
  Phys.\ Rev.\ D {\bf 78} (2008) 114033.
  doi:10.1103/PhysRevD.78.114033


\bibitem{Close:2008hv}
  F.~E.~Close,
  arXiv:0801.2646 [hep-ph].





\bibitem{Ding:2008gr}
  G.-J.~Ding,
  Phys.\ Rev.\ D {\bf 79}, 014001 (2009).

\bibitem{Dai:2012pb}
  L.~Y.~Dai, M.~Shi, G.~Y.~Tang and H.~Q.~Zheng,
  Phys.\ Rev.\ D {\bf 92}, no. 1, 014020 (2015)
  doi:10.1103/PhysRevD.92.014020
  [arXiv:1206.6911 [hep-ph]].


\bibitem{MartinezTorres:2009xb}
  A.~Martinez Torres, K.~P.~Khemchandani, D.~Gamermann and E.~Oset,
  Phys.\ Rev.\ D {\bf 80}, 094012 (2009)
  [arXiv:0906.5333 [nucl-th]].



\bibitem{Chiu:2005ey}
  T.~W.~Chiu {\it et al.} [TWQCD Collaboration],
  Phys.\ Rev.\ D {\bf 73} (2006) 094510
  doi:10.1103/PhysRevD.73.094510
  [hep-lat/0512029].

\bibitem{Liu:2012ze}
  L.~Liu {\it et al.} [Hadron Spectrum Collaboration],
  JHEP {\bf 1207}, 126 (2012)
  doi:10.1007/JHEP07(2012)126
  [arXiv:1204.5425 [hep-ph]].

\bibitem{Chen:2016ejo}
  Y.~Chen, W.~F.~Chiu, M.~Gong, L.~C.~Gui and Z.~Liu,
  Chin.\ Phys.\ C {\bf 40}, no. 8, 081002 (2016)
  doi:10.1088/1674-1137/40/8/081002
  [arXiv:1604.03401 [hep-lat]].


\bibitem{Wang:2013cya}
  Q.~Wang, C.~Hanhart and Q.~Zhao,
  Phys.\ Rev.\ Lett.\  {\bf 111}, no. 13, 132003 (2013)
  doi:10.1103/PhysRevLett.111.132003
  [arXiv:1303.6355 [hep-ph]].


\bibitem{Wang:2013kra}
  Q.~Wang, M.~Cleven, F.~-K.~Guo, C.~Hanhart, U.-G.~Mei{\ss}ner, X.~-G.~Wu and Q.~Zhao,
  Phys.\ Rev.\ D {\bf 89}, 034001 (2014)
  [arXiv:1309.4303 [hep-ph]].


\bibitem{Qin:2016spb}
  W.~Qin, S.~R.~Xue and Q.~Zhao,
  Phys.\ Rev.\ D {\bf 94}, no. 5, 054035 (2016)
  doi:10.1103/PhysRevD.94.054035
  [arXiv:1605.02407 [hep-ph]].


\bibitem{Cleven:2013mka}
  M.~Cleven, Q.~Wang, F.~K.~Guo, C.~Hanhart, U.~G.~Mei{\ss}ner and Q.~Zhao,
  Phys.\ Rev.\ D {\bf 90} (2014) 7,  074039
  [arXiv:1310.2190 [hep-ph]].

\bibitem{Guo:2013nza}
  F.~K.~Guo, C.~Hanhart, U.~G.~Mei{\ss}ner, Q.~Wang and Q.~Zhao,
  Phys.\ Lett.\ B {\bf 725} (2013) 127
  doi:10.1016/j.physletb.2013.06.053
  [arXiv:1306.3096 [hep-ph]].

\bibitem{Pakhlova:2009jv}
  G.~Pakhlova {\it et al.} [Belle Collaboration],
  Phys.\ Rev.\ D {\bf 80}, 091101 (2009)  doi:10.1103/PhysRevD.80.091101  [arXiv:0908.0231 [hep-ex]].


\bibitem{Ablikim:2014qwy}
  M.~Ablikim {\it et al.} [BESIII Collaboration],
  Phys.\ Rev.\ Lett.\  {\bf 114} (2015) no.9,  092003
  doi:10.1103/PhysRevLett.114.092003
  [arXiv:1410.6538 [hep-ex]].

\bibitem{Ablikim:2015uix}
  M.~Ablikim {\it et al.} [BESIII Collaboration],
  Phys.\ Rev.\ D {\bf 93}, no. 1, 011102 (2016)
  doi:10.1103/PhysRevD.93.011102
  [arXiv:1511.08564 [hep-ex]].


\bibitem{Li:2014jja}
  X.~Li and M.~B.~Voloshin,
  Phys.\ Rev.\ D {\bf 91} (2015) no.3,  034004
  doi:10.1103/PhysRevD.91.034004
  [arXiv:1411.2952 [hep-ph]].

\bibitem{Faccini:2014pma}
  R.~Faccini, G.~Filaci, A.~L.~Guerrieri, A.~Pilloni and A.~D.~Polosa,
  Phys.\ Rev.\ D {\bf 91} (2015) no.11,  117501
  doi:10.1103/PhysRevD.91.117501
  [arXiv:1412.7196 [hep-ph]].

\bibitem{Maiani:2014aja}
  L.~Maiani, F.~Piccinini, A.~D.~Polosa and V.~Riquer,
  Phys.\ Rev.\ D {\bf 89} (2014) 114010
  doi:10.1103/PhysRevD.89.114010
  [arXiv:1405.1551 [hep-ph]].




\bibitem{O'Connell:1995wf}
  H.~B.~O'Connell, B.~C.~Pearce, A.~W.~Thomas and A.~G.~Williams,
  Prog.\ Part.\ Nucl.\ Phys.\  {\bf 39} (1997) 201
  [hep-ph/9501251].

\bibitem{Cao:2016xqo}
  Z.~Cao, M.~Cleven, Q.~Wang and Q.~Zhao,
  Eur.\ Phys.\ J.\ C {\bf 76}, no. 11, 601 (2016)
  doi:10.1140/epjc/s10052-016-4448-3
  [arXiv:1608.07947 [hep-ph]].


\bibitem{Korner:1992pz}
  J.~G.~Korner, D.~Pirjol and K.~Schilcher,
  Phys.\ Rev.\ D {\bf 47} (1993) 3955

\bibitem{Isgur:1990jf}
  N.~Isgur and M.~B.~Wise,
  Phys.\ Rev.\ D {\bf 43} (1991) 819.
  doi:10.1103/PhysRevD.43.819


\bibitem{Ablikim:2013dyn}
  M.~Ablikim {\it et al.} [BESIII Collaboration],
  Phys.\ Rev.\ Lett.\  {\bf 112}, no. 9, 092001 (2014)
  doi:10.1103/PhysRevLett.112.092001
  [arXiv:1310.4101 [hep-ex]].


\bibitem{Meissner:1987ge}
  U.~G.~Mei{\ss}ner,
  Phys.\ Rept.\  {\bf 161} (1988) 213.
  doi:10.1016/0370-1573(88)90090-7







\bibitem{Guo:2010ak}
  F.~K.~Guo, C.~Hanhart, G.~Li, U.~G.~Meissner and Q.~Zhao,
  Phys.\ Rev.\ D {\bf 83} (2011) 034013
  doi:10.1103/PhysRevD.83.034013
  [arXiv:1008.3632 [hep-ph]].

\end{thebibliography}
\end{document}